\documentclass[12pt]{article}

\usepackage{url}

\usepackage{amsmath}
\usepackage{amssymb}
\usepackage{graphicx, hyperref}
\usepackage{cite}

\usepackage{color} 
\usepackage{colortbl}
\usepackage[final]{pdfpages}
\usepackage{setspace}

\usepackage[comma, round, sort]{natbib} 
\usepackage{lineno}

\newcommand{\F}{\mathcal{F}}

\newcommand{\Fst}{$F_{ST} \ $}

\usepackage{soul}
\usepackage[utf8]{inputenc}

\title{A New Local Score Based Method Applied to Behavior-divergent
Quail Lines Sequenced in Pools Precisely Detects Selection Signatures on Genes Related to Autism}

\author{Mar\'ia In\'es Fariello,$^{1,2,3,4,5}$
Simon Boitard,$^{6,7}$
Sabine Mercier,$^{8,9}$ \\
David Robelin,$^{1,2,3}$
Thomas Faraut,$^{1,2,3}$
C\'ecile Arnould,$^{10}$ \\
Julien Recoquillay,$^{11}$
Olivier Bouchez, $^{1,2,3,12}$
G\'erald Salin,$^{1,2,3,12}$ \\
Patrice, Dehais,$^{13}$
David Gourichon, $^{14}$
Sophie Leroux, $^{1,2,3}$
Fr\'ed\'erique Pitel,$^{1,2,3}$\\
Christine Leterrier,$^{10}$ 
Magali SanCristobal,$^{\ast,1,2,3,9,15}$ 
}

\begin{document}

\maketitle
\begin{flushleft}
{$^1$} INRA, UMR1388 G\'en\'etique, Physiologie et Syst\`emes d’Elevage, F-31326 Castanet-Tolosan, France \\
{$^2$} Universit\'e de Toulouse INPT ENSAT, UMR1388 G\'en\'etique, Physiologie et Syst\`emes d’Elevage, F-31326 Castanet-Tolosan, France \\ 
{$^3$} Universit\'e de Toulouse INPT ENVT, UMR1388 G\'en\'etique, Physiologie et Syst\`emes d’Elevage, F-31076 Toulouse, France \\
{$^4$} Unidad de Bioinformática, Institut Pasteur, Montevideo, Uruguay \\
{$^5$} Facultad de Ingeniería, Universidad de la República, Montevideo, Uruguay \\
{$^6$} UMR1313 G\'en\'etique Animale et Biologie Int\'egrative, INRA \& AgroParisTech, F-78530 Jouy-en-Josas, France \\
{$^7$} UMR7205 Origine, Structure et Evolution de la Biodiversit\'e, MNHN \& EPHE \& CNRS \& UPMC, F-75005 Paris, France\\
{$^8$} Universit\'e de Toulouse II, UFR SES, D\'epartement Math\'ematique-Informatique, 5 all\'ee Antonio Machado, F-31058 Toulouse cedex 09, France \\
{$^9$} Universit\'e de Toulouse, UMR5219, Institut de Math\'ematiques, F-31077 Toulouse, France \\
{$^{10}$} Unit\'e de Physiologie de la Reproduction et des Comportements, UMR INRA – CNRS – Universit\'e de Tours, France \\
{$^{11}$} INRA, UR83 Recherches Avicoles, F-37380 Tours, Nouzilly, France \\  
{$^{12}$} INRA, GeT-PlaGe Genotoul, F-31326 Castanet-Tolosan, France \\
{$^{13}$} INRA, SIGENAE, F-31326 Castanet-Tolosan, France \\
{$^{14}$} UE1295 Pôle d’Expérimentation Avicole de Tours, F-37380 Nouzilly, France \\
{$^{15}$} INSA, D\'epartement de G\'enie Math\'ematiques, F-31077 Toulouse cedex 4, France 
 \end{flushleft}

\begin{flushleft} 

{{\bf Running title:} Selection signatures in divergent quail lines}  \\


{{\bf Corresponding Author:}}\\ 
Magali San Cristobal\\
INRA, UMR1388 Génétique, Physiologie et Systèmes d’Elevage (GenPhySE)\\
  24 Chemin de Borde Rouge\\
	Auzeville, CS 52627 \\
F-31326 Castanet Tolosan Cedex,\\
France\\
(+33)561285122\\
E-mail: magali.san-cristobal@toulouse.inra.fr

\end{flushleft}

\abstract{Detecting genomic footprints of selection is an important step in the understanding 
of evolution. Accounting for linkage disequilibrium in genome scans  
allows increasing the detection power, but haplotype-based methods require individual 
genotypes and are not applicable on pool-sequenced samples. 
We propose to take advantage of the local score approach to account for linkage disequilibrium, accumulating 
(possibly small) signals from single markers over a genomic segment, 
to clearly pinpoint a selection signal, avoiding windowing methods. 
This method provided results similar to haplotype-based methods on two benchmark data sets with individual genotypes.
Results obtained for a divergent selection experiment on behavior in quail, where two lines were sequenced in pools,
are precise and biologically coherent, while competing methods failed:
our approach led to the detection of signals involving genes known to act on social responsiveness or autistic traits. 
This local score approach is general and can be applied to other genome-wide analyzes such as GWAS 
or genome scans for selection.}

\newpage

\doublespacing

\section{Introduction} 
\label{sec:Intro}

Detecting genomic regions that have evolved under selection is a long standing question in population genetics,
which has received a growing interest these last years.
Due to linkage disequilibrium (LD), a selection signature from a positive 
selected polymorphism is not limited to the causal polymorphism,
but generally is extended to a wider genomic interval including it. 
Markers in the neighborhood of the selection target might thus also show departures 
from the expected patterns under the neutral evolution null hypothesis \citep{PRIPIC2010}.
In addition, several positively selected polymorphisms might be found in one gene.
Consequently, detection power is expected to increase when searching for regions with 
outlier genetic diversity patterns, rather than considering markers independently 
from each other.
Several detection methods taking advantage of LD have been proposed, testing a single population
\citep{EHH,iHS,BOI2009,NIE2005}; pairs of populations 
\citep{XP-EHH,BHA2011,CHE2010} or a large number of populations simultaneously,
while accounting for the hierarchical structure of these populations (hapFLK 
\citet{FAR2013}). In simulation scenarios with 
one single site under positive selection, 
the latter method outperformed single marker tests such as 
$F_{ST}$ \citep{WC84} or $\F$-LK \citep{Maxime}, as well as their windowed versions 
\citep{WEI2005}, except for sequencing data where all polymorphic sites 
(including the one under selection) were observed. It also provided increased detection power when 
compared to a cross population haplotypic
tests \citep{XP-EHH} in most scenarios, specially when the genetic drift was high.

Haplotypic methods \citep{EHH,iHS,XP-EHH,FAR2013} require genetic data at the individual level 
and are rather computationally demanding. 
On the other hand, single marker statistics have a lot of variability \citep{WEI2005} 
and high values of these statistics can be reached just because of genetic drift.
When the expected genetic drift of a population is high, the variance of allele frequencies 
is expected to be large \citep{Maxime}, so the probability of false positives when using 
single markers tests is particularly high. To take advantage from the linkage disequilibrium information
while using genetic data at the population level, an alternative approach is to compute single marker statistics
and combine them locally.

Following this approach, \cite{WEI2005} proposed to average single 
marker $F_{ST}$ within sliding windows along the genome, in order to detect regions with outstanding differentiation 
between populations, while smoothing the profile of the statistic genome wide.  
Sliding windows approaches imply to choose a window size, which is usually done arbitrarily. To overcome this problem
other strategies to find clusters of high $F_{ST}$ values were proposed 
\citep{MYL08,JOH2010}. 
\cite{MYL08} proposed an algorithm to find clusters of
markers with $F_{ST}$ values in the top $1 \%$ of the genome-wide distribution 
(called the $T1$ SNPs). For each SNP $t_i \in T1$, the authors counted the number of 
non$-T1$ SNPs located between $t_i$ and $t_{i+9}$. The resulting number
was defined as the clustering coefficient $K$. Low values of $K$ should 
point regions of high density of $T1$ SNPs.
\cite{JOH2010} proposed a different algorithm in the same spirit. 
The authors considered that two SNPs were in the same cluster if the distance between them 
was shorter than 1Mb. Here, instead of taking the top $1 \%$ of
a distribution, they considered SNPs that were differentially fixed, \emph{i.e.} with 
one allele lost in one population and fixed in the other one. 
Although the two above approaches avoid to define fixed windows, they involve tuning 
parameters whose value is generally fixed arbitrarily. 
Windows with 5, 10, 15 and 20 $T1$ SNPs were considered by \cite{MYL08}, and  
regions with 2 or 5 SNPs in a Mb by \cite{JOH2010}, but the authors
acknowledged that this parameter choice was subjective.
The same occurred with the creeping window strategy in \citet{Qanb2012}, which is suited for testing one line at a time.

The objective of this study is to propose a new approach based on the local score theory, 
to detect genomic regions under selection by the combination of single marker tests.
As the methods described above, we focus here on the detection of regions with outstanding differentiation 
between populations, which can be measured by single marker statistics as $F_{ST}$ or $\F$-LK.
The idea of the local score approach is to cumulate selection signals via small $p$-values of single 
markers tests (or equivalently large values of $-\log_{10}(p-value)$, further defined 
as a score) in an automatic manner. This approach is suited to the case where only allele frequencies are available, and runs much faster than haplotypic methods. In additon, it helps single marker statistics to gain power, specially in cases where 
genetic drift is high and they completely loose their power of detection.
Similar to the methods described above, it implies a tuning parameter $\xi$ (the $p$-value threshold on single marker tests), 
but this parameter has an intuitive interpretation and can be adjusted to detect different kinds of selection events.

Another advantage of the local score over existing cumulating approaches is that it is based on a solid statistical basis.
Its asymptotic distribution is known \citep{KAR1992} and an exact distribution can be computed for not too long 
sequences ($L \leq 10,000$), because of computational constraints \citep{MER2001}.
These results assume that markers are independent, but we propose here approximations that account for the correlation between markers.  
The local score approach plays an important role in bioinformatics
- it is mostly used for sequence alignment- and was also used in previous epidemiology 
\citep{GUE2006} or genome wide  association studies \citep{TEY2012} with a score 
function related to $-\log_{10}(p-value)$ as in the present study.  

We show here that the local score approach performs very well for the detection 
of selection signatures, by comparing it with the haplotype based approach hapFLK 
on 2 benchmark data sets.
We then apply it to detect genes associated to social reinstatement behavior in quail, 
using pooled NGS data from two quail lines that have been divergently selected 
for this trait.

\section{Material and Methods} 

\label{sec:NewApproaches}
\label{sec:MatMeth}		

\subsection{Data}
\label{subsec:Data}

\paragraph{Quail data}

Two divergent lines produced and maintained at the INRA experimental unit 1295 (UE PEAT, F-37380 Nouzilly, 
France) were used in the experiment. These lines with high social reinstatement behavior (HSR) and low social 
reinstatement behavior (LSR) have been divergently selected on their propensity to rejoin a group of conspecifics 
when 10-day old \cite{MIL1991}. They differed consistently on several aspects of their social behavior \citep{JON1999,RIC2008} 
and also notably on the characteristics of the social bond they developed \citep{SCH2010,SCH2011}.

A total of 10 individuals from generation 50 of each quail line were used: 3 males and 7 females, chosen as unrelated as possible. 
Genomic DNA was obtained from blood samples of these 20 animals through a high-salt extraction method \citep{ROU2003}. 
Sequencing was performed on 1 DNA pool per line, consisting of an equimolar mix of the ten samples. 
Two libraries, one for each pool, with an insert size of 300 bp, were prepared following Illumina instructions for 
genomic DNA sequencing (TruSeq DNA sample v2). 
Samples were then sequenced (paired-ends, 100 bp) on a HiSeq 2000 sequencer (Illumina), by using one lane per line (TruSeq SBS kit v3).

In the absence of an available genome sequence for the quail, the reads of the two divergent lines (190,159,084 and 
230,805,732 reads respectively) were mapped to the chicken genome assembly (GallusWU2.58). 
To achieve good sensitivity, the reads were aligned using the glint aligner (http://lipm-bioinfo.toulouse.inra.fr/download/) 
with default parameters. 
The glint program, a general purpose nucleic sequence aligner, was designed specifically to align medium-divergent sequences, 
characteristic of inter-specific genome comparison. 
A total of 54.6\% and 55.4\% of reads were aligned in a proper 
pair to the chicken genome (mapping quality of at least 20) for the HSR 
and LSR lines respectively, corresponding to 8 and 10X genome coverage. 
In contrast, the bwa aligner \citep{LI2009} was only able to align 10.2\% and 10.1\% of the reads respectively (less than 2X coverage). 
The alignments were first converted into the pileup format using the mpileup command of samtools with options -B, -q 20 and -f.
Within each line, the frequency of the reference chicken allele was estimated for all SNPs that were covered by at least 5 reads, 
using Pool-HMM \citep{BOI2013} with the options --estim, --a reference and -c 5.
Pool-HMM accounts for the sampling effects and the sequencing error probabilities inherent to pooled NGS experiments, 
when estimating allele frequencies.
$\F$-LK values were finally computed at all SNPs for which allele frequency data had been obtained in the two lines, using 
private python scripts.

\subsection{Local score approach} 
We propose to highlight segments of adjacent loci with small $p$-values, 
starting from results of single marker tests along the genome. Our strategy is to find 
high values of the "score" $-log10(p-value)$ (up to an additive constant) and to 
cumulate them over adjacent loci.

Let's assume that we observe data at $L$ consecutive positions along a sequence, 
and that we have chosen a score function that transforms data at position $\ell$  
into a real number $X_\ell$. Here, we chose $X_\ell = - log10(p-value_\ell) - \xi$, 
where $p-value_\ell$ is the $p$-value of a test at position $\ell$ and $\xi$ 
is an arbitrarily chosen real number. Then $X = (X_1, \dots, X_\ell, \dots, X_L)$ is 
the resulting sequence of scores. 
We consider each chromosome as a sequence and $\ell$ refers to a fixed position on the chromosome 
(for instance a SNP). 
We further introduce the Lindley process $h=(h_1,\dots, h_\ell, \dots, h_L)$, with 
$h_\ell = max (0, h_{\ell-1}+X_\ell)$ and $h_0=0$.

The local score of the sequence $X$ is defined as:
\begin{equation}
H_L (X) = \max\limits_{1 \leq \ell \leq L} h_\ell.
\label{eq:def_SL}
\end{equation}

An equivalent way to apprehend the local score is to compute the maximum of the 
partial sums of the scores at adjacent loci:
 $H_L (X) = \max \limits_{1 \leq i \leq j \leq L} \{ X_i + X_{i+1} \dots + X_j \}$.

The local score has an associated interval of interest $[\ell_{start}(X), \ell_{stop}(X)]$ 
enriched in high values of $X$. 
The end of the interval is the locus where the local score is realized, 
\emph{i.e.} $\ell_{stop}(X) = argmax_{\ell} \ h_\ell$. The interval begins on the last 
locus before $\ell_{stop}(X)$ where the Lindley process is equal to 0, \emph{i.e.} 
$\ell_{start}(X) = \max \{ \ell \le \ell_{stop}, h_\ell = 0 \} $.

\begin{figure}[!h]
\includegraphics[width=\linewidth]{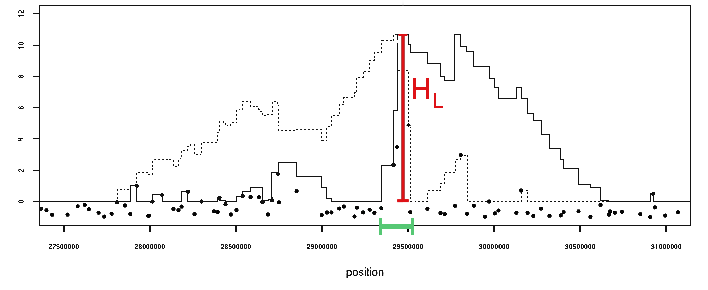}
\caption{Example of local score on a chromosome segment in sheep: region related to the presence/absence of horns.
Single marker significance is displayed by black points, representing $-log10(p-value_{\F-LK})-1$ for a $\F$-LK test of neutrality. 
Three consecutive SNPs with high scores have been associated to the presence/absence of horns \citep{SLA2010}.
The Lindley process of the score function $-log10(p-value_{\F-LK}) -1$ is drawn with a solid line going from the left to the 
right of the chromosome, and with a dashed line from the right to the left. The local score $H_L$  is achieved at the red 
arrow. The corresponding segment, materialized in green, contains exactly the 3 SNPs.}
\label{fig:horns}
\end{figure}

The interval of interest is unchanged if we read the sequence in the opposite direction: 
denoting $\bar X = (X_L, \dots, X_\ell, \dots, X_1)$ (\emph{i.e.} 
$X_i =\bar{X}_{L-i+1}$), it can be shown that $\ell_{start}(X) =  \ell_{stop}(\bar X)$ 
and $\ell_{stop}(X) = \ell_{start}(\bar X)$, so the interval realizing the local score is easily highlighted.
The notions of Lindley process, local score and the associated interval are illustrated in 
Figure \ref{fig:horns}.

The expectation $E(X)$ of the score should be negative to ensure the Lindley process to vanish at least once,
in order to focus on local, instead of global, interest. 
Assuming that $p-value_\ell$ follows a uniform distribution, then $E(X) = \frac{1}{log(10)}-\xi$, 
so $\xi$ should be approximately greater than $0.43$.
On the other hand, at least one score should be positive ($\max \limits_{\ell} X_\ell >0$), which implies that 
$\xi < - log10( \min \limits_{\ell}(p-value_\ell) )$.
In practice, it is up to the user to choose a value between $\frac{1}{L} \sum \limits_{\ell} -log10 (p-value_\ell)$ 
(close to 0.43 if the $p$-values are approximately uniform under the null), 
and $- log10( \min \limits_{\ell}(p-value_\ell) )$. 
The greater $\xi$, the closer we are to the single marker approach, since only the top markers will 
contribute to the Lindley process.
The lower $\xi$, the more often local maxima will appear in the Lindley process, and the larger will 
be the detected region.

\paragraph{Distribution of the local score under the null hypothesis}
For independent sites (i.i.d. model) and integer scores it is possible to obtain exact $p$-values for short 
sequences and approximate $p$-values for long sequences \citep{KAR1990,KAR1992,MER2003}. Exact $p$-values 
can be obtained in Markovian cases \citep{HASS2007} but with a large computational time for such sequence 
length.

Real data however hardly meets the independence assumption. In quail data, the correlation of score function values of adjacent 
sites ranged from 0.22 to 0.47 for the 28 quail chromosomes, with a vast majority around 0.3.
The maximum value (0.47) was reached for the "shortest" chromosome, GGA16, which encountered alignment problems and was thus 
not included in further analyzes.

For a non zero correlation between sites, we propose to obtain approximate $p$-values for 
the local score $H_L$ on a chromosome of length $L$, assuming a constant auto-correlation 
along the genome. 
A first possibility is to compute an empirical distribution of the local score $H_L$ under 
the null hypothesis of neutrality,
by way of simulations involving a given auto-correlation $\rho$ for the score function.
This was achievable in quail just for short chromosomes (\emph{i.e.} GGA28,  
40{,}000 bp long), but the computational time was prohibitive for the largest ones 
(\emph{i.e.} GGA1, 2{,}631{,}000 bp long).

Another possibility is to follow ideas of \cite{GUE2006} and \cite{ROB2005},
and to approximate the distribution of the local score with a Gumbel distribution, which in general is used to model
maxima of stochastic processes. As the cumulative density function of a Gumbel variable is linear after 
a log(-log) transformation, for given values of the sequence length $L$ and the correlation $\rho$, an empirical 
cumulative density function (ecdf) 
$F_{L,\rho}$ can be obtained by simulating sequences with these properties and
fitting a linear model of the form
\begin{equation}
 log(-log(F_{L,\rho}(y)) = a_{L,\rho} + b_{L,\rho} y + e,  
\label{eq:ecdfGumbel}
\end{equation}

\noindent with $e \sim {\cal N}(0, \sigma_{L,\rho}^2)$, 
which provides estimates $\hat a_{L,\rho}$ and $\hat b_{L,\rho}$ of the parameters 
$a_{L,\rho}$ and $b_{L,\rho}$ (see Figure S2).

We applied this strategy for several values of $L$ (from 100 to 45000)   
and $\rho$ (from 0 to 0.9) (Figure S2). 
For each pair $(L,\rho)$   
 5000 sequences $z_{i=1,...,L}$ were generated from a 
multivariate uniform distribution with autocorrelation $\rho$, mimicking the null distribution of 
$p$-values of single marker tests.
Details of these simulations are given in Supporting Information, Section: 
Null distribution of the local score for correlated markers. 

For a given value of $\xi$, abacuses of parameters $\hat a_{L,\rho} - log(L)$ and $\hat b_{L,\rho}$  
converge rapidely to a limit distribution from 
approximately $L=10{,}000$ (Figure S3). 
Thus, if the $p$-values follow a uniform distribution, 
for sufficiently long sequences $a-log(L)$ and $b$ depend only of $\rho$
and are independent from $L$. Taking advantage of this property,
we further implemented a cubic and a quadratic fit
that provide general approximations $\tilde \hat a_{L,\rho}$ and $\tilde \hat b_{L,rho}$ of $\hat a_{L,\rho}$ 
and $\hat b_{L,\rho}$
for any value of $L$ and $\rho$ (not only those simulated).
These formulas can be found in Supporting Information, for $\xi=1$ and 2, and a script is given to compute the polynomes for 
other values of $\xi$.
The $p$-value for the local 
score of any chromosome can be computed by first obtaining $a$ and $b$ from those formulas and then using equation 
(\ref{eq:ecdfGumbel}), that is:
\begin{equation}
 P( H_L \leq x) \approx 1 - \exp \left( - \exp ( \tilde \hat a_{L,\rho} + \tilde \hat b_{L,\rho} x) \right)
\label{eq:pvalGumbel}
\end{equation}

\noindent Remark that this formula has the same form as Karlin's approximation in the independent case
($\rho=0$), see Supporting Information.
Similarly, threshold values $t_{L, \rho ; \alpha}$ at $\alpha$ 
level for the local score can be obtained by 
\begin{equation}
 t_{L, \rho ; \alpha} \approx \frac{ log ( -log (1-\alpha) )- \tilde \hat a_{L,\rho} } { \tilde \hat b_{L,\rho}  }.
\label{eq:thresGumbel}
\end{equation}

\noindent This strategy was applied for the analysis of the sheep data, and could more generally be used by any 
future study for which the $p$-values of single marker tests
are uniformly distributed under the null hypothesis.

However, as described in the Data Section (\ref{subsec:Data}) the quail data has many particularities. One of 
them is a particularly strong genetic drift, resulting in a very high frequency of fixed alleles.
This implies that the neutral distribution of $\F$-LK is no longer a chi-square, in contrast to what is assumed when 
computing the $p$-values of this test.
Consequently, the $p$-value distribution of $\F$-LK under the null hypothesis is not uniform and the strategy described 
above cannot be applied in this particular case.

To get an approximate null distribution for the local scores obtained from the quail data, we thus used a slightly different 
strategy,
based on the re-sampling of score sequences of fixed length from the quail data (see the Supporting Information for more details,
Figures S4-S6, Tables S1-S2). 
This provided estimates
$\tilde \hat a_{L,\rho}$ and $\tilde \hat b_{L,\rho}$ 
of $\hat a_{L,\rho}$ and $\hat b_{L,\rho}$, and hence of $ a_{L,\rho}$ and $ 
b_{L,\rho}$ :
\begin{eqnarray}
\hat a_{L,\rho} =   log(L) -7.60-6.86 \rho \\
\hat b_{L,\rho} =  -0.49 +1.51 \rho 
\end{eqnarray}

\noindent which are only valid locally around $\rho=0.3$, and are thus applicable to all quail chromosomes except 
 GGA16 (whose auto-correlation is 0.47).

\section{Results} 
\label{sec:Results}

\subsection{Benchmark 1: Lactase region of HapMap data}
We tested a 4Mb region (134-138 Mb) on Human chromosome 2 containing the \emph{LCT} gene, 
because a known causal mutation for the lactase 
persistent phenotype in the CEU (Utah Residents with Northern and
Western European ancestry) population is located in chromosome 2 at position 136,325,116.     
Data was taken from the HapMap Phase III dataset and consisted in the genotypes of 370 founder 
individuals from the CEU, TSI (Toscani in Italy), CHB (Han Chinese in Bejing, China) and JPT (Japanese in Tokyo, Japan) populations. 
Only $25 \%$ of the available SNPs (that is 497 SNPs) 
were included in the analysis.

\begin{figure}[!h]
\includegraphics[width=\linewidth]{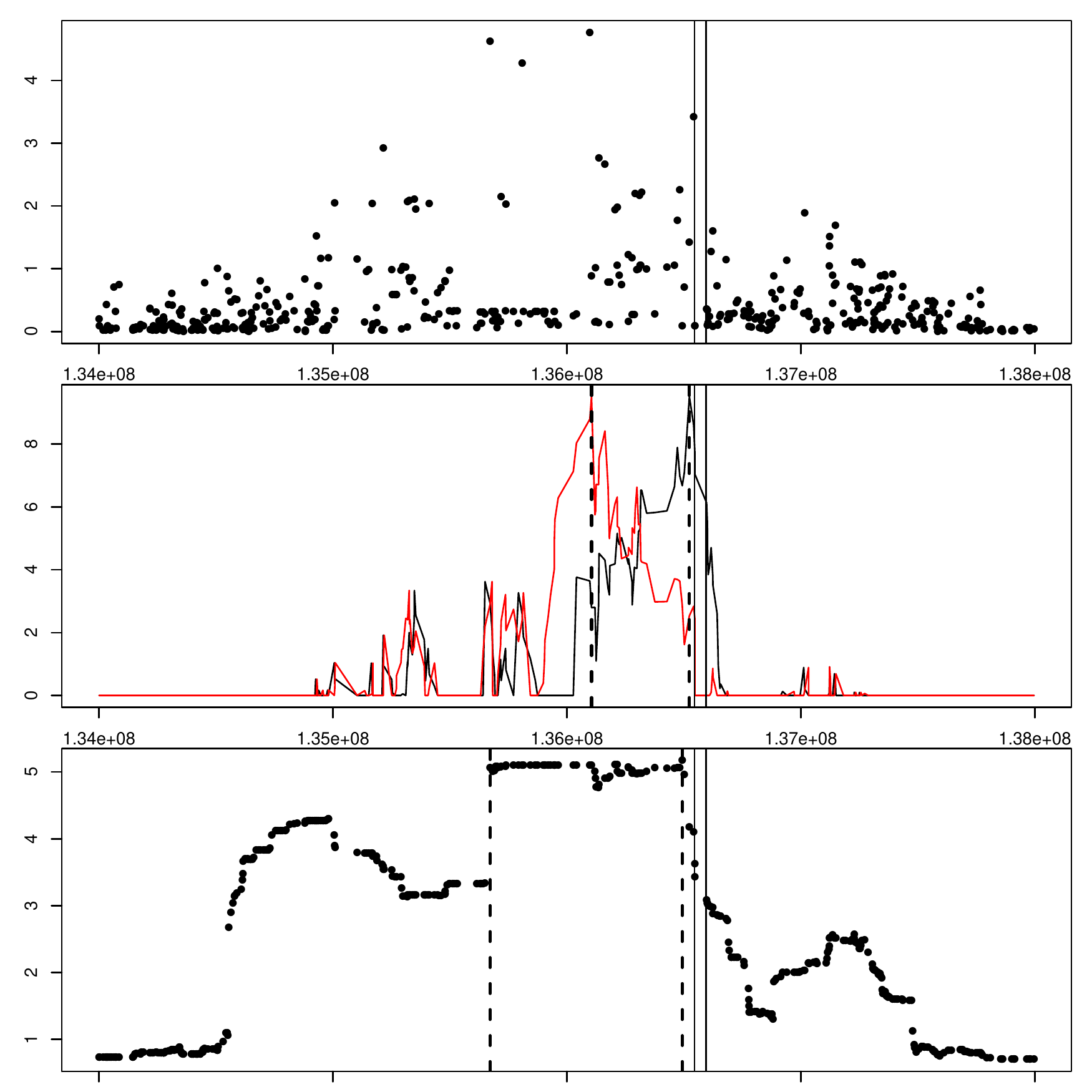}
\caption{Selection footprints in human HapMap data: focus on Lactase region. 
{\bf{(Top)}} $-log10(p-value_{\F-LK})$ of the single marker $\F$-LK test. 
{\bf{(Middle)}} Lindley process based on the score function $-log10(p-value_{\F-LK})-1$ starting from the left (black) 
and from the right (red). 
Dotted vertical lines indicate the limits of the detected interval (achieving the local score).
{\bf{(Bottom)}} hapFLK values. Dotted vertical lines indicate the limits of the detected interval.
The Lactase gene is located within the 2 vertical solid lines.}
\label{fig:lactase}
\end{figure}

We compared results obtained from the single locus approach $\F$-LK \citep{Maxime},
the local score and the haplotypic approach hapFLK \citep{FAR2013} (Figure \ref{fig:lactase}). 
While the top markers for $\F$-LK were quite far from the Lactase gene, 
The intervals given by hapFLK and the local score were really close to the Lactase 
gene, with the local score providing a smaller interval. 
Here, the local score was based on the score function $-log10(p-value_{\F-LK})-1$,
meaning that $p$-values of single marker tests greater than $10^{-1}$ 
were cumulated to find an interval achieving the local score.
On this benchmark region, the local score clearly highlighted a well known target 
of selection in Human, thus performing as well as the hapFLK test.

\subsection{Benchmark 2: Sheep HapMap data}
The Sheep HapMap dataset includes individual genotypes at 60K SNPs for 2819 animals from 74 worldwide 
sheep breeds \citep{kijas-etal-12}.
A genome scan for selection was performed on this dataset using single marker \Fst \citep{kijas-etal-12}.
We considered here a subset of this dataset: a group of
4 breeds (two Spanish breeds and two French breeds) originating from South-western Europe (256 sheep).
Genome scans for selection using the single marker test $\F$-LK and the haplotypic test hapFLK have already 
been performed in this group \citep{FAR2014}. Interestingly, genome scans with $\F$-LK and hapFLK lead to 
distinct detected regions.

Here, the score was $X_\ell = - log10(p-value_{\F-LK})-\xi$, with $\xi=1$ or 2 (Figure S7). 
The approaches taking account of the dependence between adjacent markers (the local score and hapFLK) gave 
a clear picture of different selection signatures. Using the single SNP test $\F$-LK, with a false discovery 
rate of 5\%, none of SNPs reached the significance threshold.

The local score approach highlighted a region on OAR10 (either for $\xi=1$ or 2) that was 
not significant either for $\F$-LK or for hapFLK (Figure \ref{fig:horns}). 
In this region, no particular haplotype was selected, but a few SNPs whose alleles were shared in several haplotypic 
backgrounds showed a moderate signal. This region is very close to \emph{RXFP2} 
for which polymorphisms have been shown to affect horn size and polledness in the Soay \citep{SLA2010} and 
Australian Merino \citep{dominik2012}.

The CPU time for a whole genome scan with hapFLK was about a few hours, while it took 
only a few minutes to compute $\F$-LK and a few more minutes to compute the local score. The significance 
threshold for each chromosome was computed using Equations (S9) and (S11) in 
Supporting information.
This computation time does not include the simulations that were used to obtain the $p$-values,
because these simulations had to be performed only once and the formulas we obtained can now be used directly,
if the $p$-values follow a uniform distribution.

\subsection{Quail data}
We looked for genome wide selection signatures  in two divergent Quail lines.   
We used the local score associated to the score function $-log10(p-value_{F_{ST}})-1$, 
($F_{ST}$ and $\F$-LK are equivalent here).

\begin{figure}[!h]
\includegraphics[width=1\linewidth]{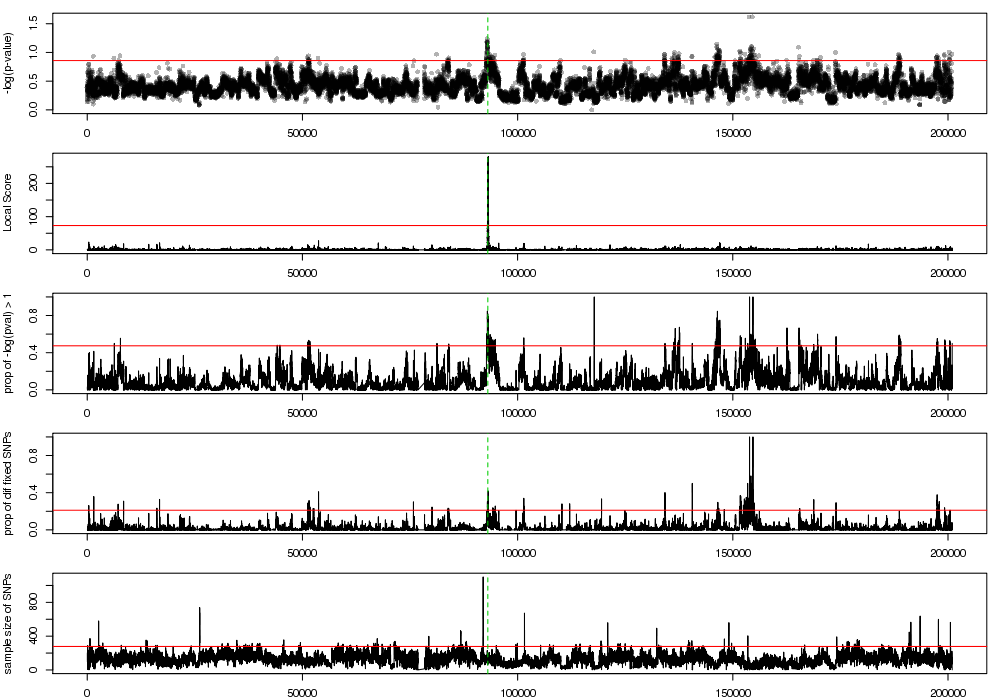}
\caption[Selection footprints in quail data, focus on GGA1]{Selection footprints in quail data, focus on GGA1. 
From top to bottom: $-log10(p-value)$ for windowed $F_{ST}$ (mean $F_{ST}$ on windows of 10kb), 
local score (Lindley process) of score function $-log10(p-value_{F_{ST}})-1$,
proportion of SNPs with a $F_{ST}$ $p$-value greater than 0.1 (windows of 10kb), 
proportion of fixed SNPs in each window, and number of SNPs in the window. Positions are indicated in Mb.
Red lines indicate the thresholds corresponding to the top $1 \%$ of the genome for each measure, except for the local score's  threshold
which has been computed for each chromosome (Supporting Information). Green lines indicate the 
center of the region detected by the local score approach.}
\label{fig:caille1_4stats}
\end{figure}

We first focused on GGA1 and compared the local score approach with 
several sliding window statistics (Figure \ref{fig:caille1_4stats}): 
(i) the mean $F_{ST}$, (ii) the proportion of "significant" SNPs, 
i.e. such that $-log10(p-value_{F_{ST}})$ is greater than 1, and (iii) the proportion of differentially 
fixed SNPs, i.e. that have reached fixation in one line and were lost in the other one.
These statistics were computed over windows of 10kb, which included 132 SNPs on average.
We used the top $1 \%$ of the distribution of the statistic genome wide
as significance threshold for this windowed statistics.
All statistics had a local maximum in the same region as the Lindley process, indicating that in this 
particular region there is an excess of differentially fixed SNPs. On the other hand, the large number of local 
maxima exceeding the threshold for these statistics, suggests that many of them may be false positives. 
A second peak, at position roughly 150000, was observed with almost all statistics but not with the local score.

\begin{figure}[!h]
\includegraphics[width=\linewidth]{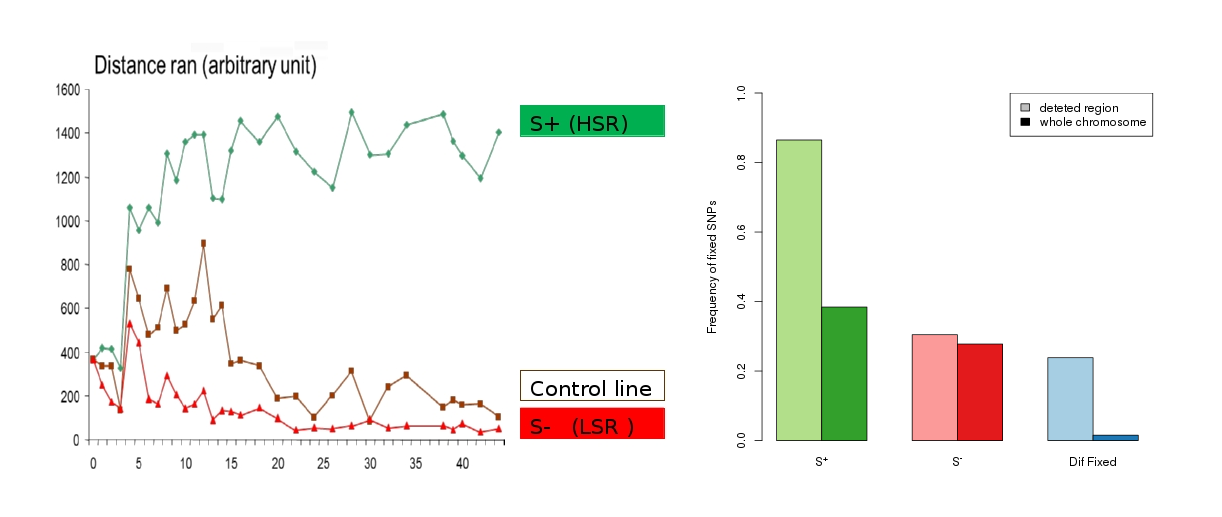}
\includegraphics[width=0.6\linewidth]{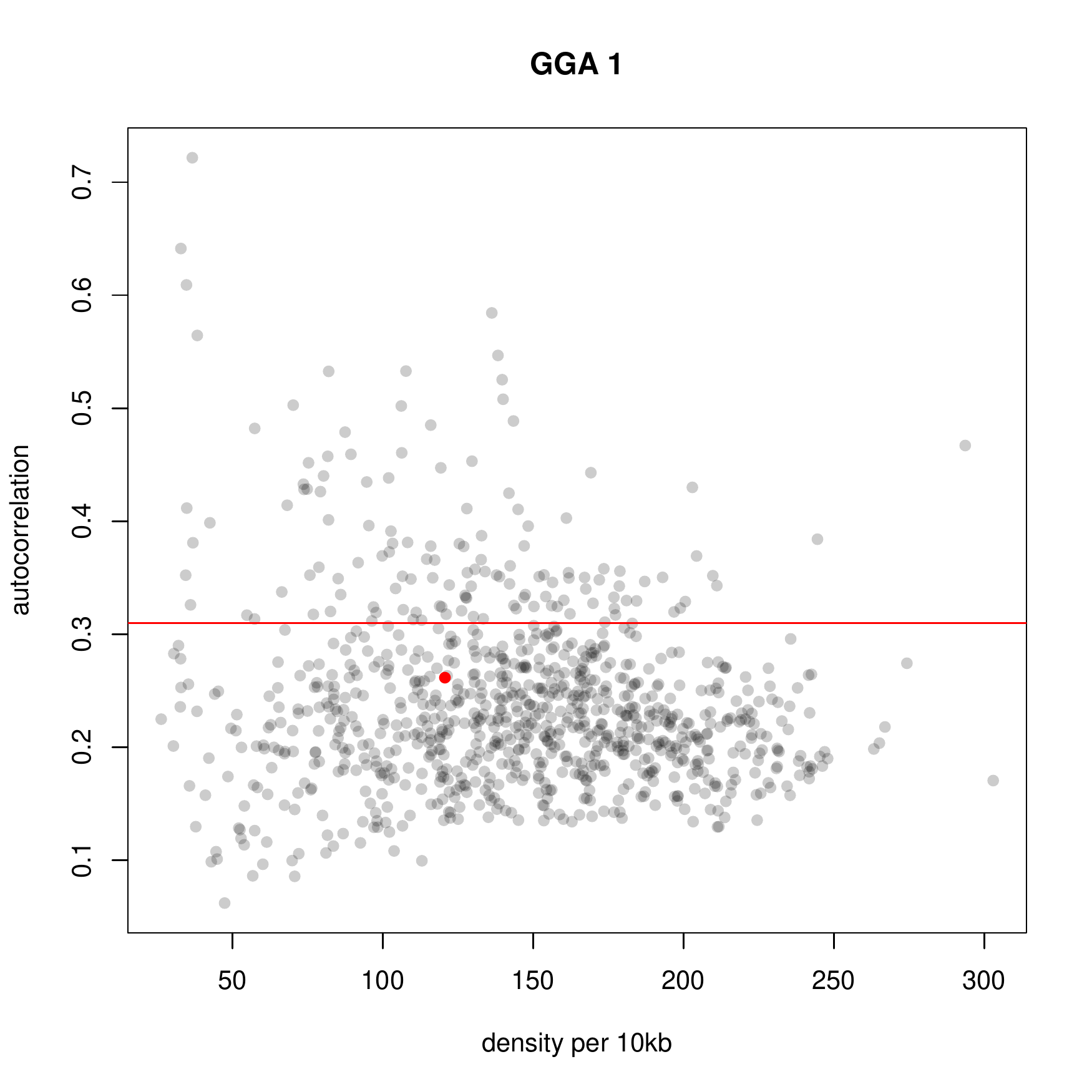}
\caption{Quail data. Top left: Evolution of the mean run distance (in an arbitrary unit) as a 
function of the generation number of the selection experiment. 
Top right: Barplot of the proportion of fixed SNPs in each line, in GGA1 (dark) and in the region 
detected by the local score approach
(light). The third pair of bars corresponds to the proportion of SNPs fixed for distinct alleles in the two lines. 
Bottom: Autocorrelations and SNP densities computed on windows of the 
same length as the detected region. The red point corresponds to the
values of the detected region. The red line represents the autocorrelation of whole chromosome 1.}
\label{fig:AutoCor-DensiteGGA1}
\end{figure}

The region detected by the local score on GGA1 was 219.23 kb long and contained 2646 SNPs. 
The SNP density in this region was 74.8 SNPs per 10kb. It is less than the whole genome 
average, suggesting that this region is probably not a false positive due to a greater density of SNPs.
In addition, the autocorrelation in this region was not an outlier,
compared to the autocorrelations of windows of the same size in the same chromosome (Figure \ref{fig:AutoCor-DensiteGGA1}).
The region displayed a slightly decreased average $\F$-LK p-value, and a slightly larger proportion 
of "significant" SNPs or fixed SNPs per window (Figure \ref{fig:caille1_4stats}), 
but this did not appear very clearly on the graphs, in contrast with the local score signal.

The local score (maximum value of the Lindley process) was in the range 11-50 for almost 
all chromosomes, except chromosomes 
GGA1 (with a local score equal to 279.73), GGA2 (local score 150.78), GGA3 (local score 184.25) 
and GGA6 (local score 286.43) (Table S3, Figure S10). 
A very clear peak was observed in these cases, in contrast with windowed $F_{ST}$ 
(Figures \ref{fig:caille1_4stats}, S8  and S9). 

Detected regions, mapped on the chicken genomic sequence, are listed in 
Table \ref{tab:caille_liste_regions}. Local score profiles and the $5 \%$ and $1 \%$ rejection thresholds 
are included in the Supplementary Information. Detected regions were in general very short 
(from 19 to 219kb) and included at most 3 genes. 
This accurate detection of the region under selection is a clear advantage for understanding the 
selective process that has been acting in these lines,
and holds great promises for the identification of the exact polymorphisms under selection, 
which could be the aim of subsequent studies.
The two regions of chromosome 4 do not reach the $5 \%$ threshold, but we decided to include them
because they contain interesting candidate genes (see Discussion).

\section{Discussion} 
\label{sec:Discussion}

The work presented here showed a clear added value of a statistical method and a data set.
The methodology developed was the only one to provide a short list of candidate genes related to the selected behavioral trait in the quail lines.

\paragraph{Candidate genes} 
Several genes comprised or partially localized in the detected regions 
under selection in Quail, have been associated with autistic disorders 
(\url{http://genome.ucsc.edu/cgi-bin/hgTracks}). 
\emph{PTPRE} (receptor-type tyrosine-protein phosphatase epsilon precursor, in region 7) is one of the 
candidate genes present on human chromosome 10q26 and has been  shown to be involved in autism spectrum 
disorder \citep{TON2011}. Similarly, \emph{ARL13B}  (in region 1) is one of the genes involved in the Joubert 
syndrome \citep{CAN2008}, a psychiatric disorder with possible autistic symptoms \citep{DOH2009}.
Finally, \emph{IMPK} (inositol polyphosphate multikinase, in region 8) maps to a position homologous 
to a region of human chromosome 10 (10q21.1) which shows a male-only signal of linkage with a social 
responsiveness trait \citep{DUV2007}. 
This study detected social responsiveness quantitative trait loci in multiplex autism families. As the 
linkage region in human is quite large, 
more work has to be done before considering this gene as a functional candidate.
 
Autistic spectrum disorders are observed in a number of disorders that have very different etiology, including 
fragile X Syndrome, Rett Syndrome or Foetal Anticonvulsant Syndrome. While these disorders have very different 
underlying etiologies, they share common qualitative behavioral abnormalities in domains particularly relevant 
for social behaviors such as language, communication and social interaction \citep{RUT1978, AME2000}. In line 
with this, a number of experiments conducted on High Social Reinstatement (HSR) 
and Low Social Reinstatement (LSR) quails indicate that the selection program carried out with these lines is 
not limited to selection on a single response, social reinstatement, but affect more generally the ability of 
the quail to process social information. 
Differences in social motivation, but also individual recognition have been described between LSR and HSR quail 
\citep{FRA1999, SCH2010}.  Inter-individual distances are longer  in LSR quail \citep{FRA1999} and LSR young quails 
have decreased interest in unfamiliar birds \citep{FRA2000} and lower isolation distress than HSR ones 
(see \cite{JON1999} for review). A last interesting candidate gene is \emph{CTNNA2} (catenin alpha 2, in region 
6). This gene, involved in synaptic plasticity, has been shown to be implicated in several behavioral traits, 
and has recently been associated with excitement-seeking, partly related to social behavior \citep{TER2011}. 
These genes may thus represent particularly relevant candidates to explain the difference between two quail 
lines that diverge on many aspects of social behavior. Further experiments will be required to examine the 
possible functional link between the selected genes and the divergent phenotype observed in these lines.

\paragraph{Generality of the local score}

We proposed a new method to detect selection signatures genome-wide, which takes linkage disequilibrium into 
account even 
when individual genetic information is not available. The strategy is to use the local score theory (which is 
widely used in sequence alignment, for example) 
to cumulate information of single marker significance over the genome, in order to detect intervals of high 
cumulated significance. 

This strategy only requires a significance or importance value (the score) for each marker.
Here we focused on the detection of regions with outstanding genetic differentiation between 
populations and proposed to build the score from the $p$-value of a single marker test (\emph{e.g.} $F_{ST}$ 
or $\F$-LK). 
Thus, the distribution of the local score only depends on the distribution of single marker test $p$-values,
which, under the null hypothesis, can generally be assumed to be uniform. 
This approach gave a clear picture of the selection targets in 2 benchmark data sets (the human HapMap and the 
SheepHapMap) including individual genotypic information obtained from SNP chips. 
In SheepHapMap data we detected signals that were not detected by hapFLK (\emph{e.g.} the horn SNPs). 
We also failed to recover some of the signals detected by hapFLK, but this may be related to the low SNP 
density of this data set. 
Indeed, as the local score approach is based on the concentration of SNPs with low $p$-values,
even strong selection events might be difficult to detect if only few SNPs are found in the region.

With high density data, the local score might thus detect a wide range of signals, from moderate signals 
spanning over a large region to strong but short signals.
The former class of signals could typically arise from recent selection events where the selected haplotype 
did not yet fix in the population (this is the case for the human Lactase signal),
as well as from recent soft sweeps, i.e. where the selected allele is associated to several haplotype backgrounds.
In contrast, the latter class of signals would correspond to old hard sweep signals, whose length has already 
been reduced by recombination. This is for instance the case of the horn locus in sheep.

\paragraph{Choosing $\xi$}  

Among other factors, the type of selection events that are detected with the local score approach depends on 
the choice of $\xi$, the threshold below which small $p$-values are cumulated.
This value must be included between 0.43 and $-log10( \min \limits_{\ell}(p-value_\ell) )$, with high values 
leading to focus on the few strongest SNPs and low values allowing to cumulate a larger number of
intermediate signals. In the quail data, the large amount of drift implies that the best single marker 
$p$-value (corresponding to an allele frequency of 0 in one line and 1 in the other) was $10^{-1.5}$,
so the upper bound for $\xi$ was 1.5.
In this case, cumulating $p$-values under $10^{-1}$ appeared appropriate, since it allowed to cumulate a 
large number of signals, while excluding $p$-values above $10^{-1}$ that can hardly be considered as significant.
Besides, due to the short evolution time since the divergence of the two lines (and consequently the advent of 
the selective constraint in each of them), 
we expected to detect recent soft or incomplete sweeps rather than old hard sweeps, which again argues for the 
use of a low value of $\xi$.

In the sheep data (Figure S7), 
the use of two different values of $\xi$ seemed to lead to two 
different detection patterns:
for $\xi=1$ the data showed more local maxima than for $\xi=2$. But, when thresholds were computed, we observed 
that the significant local maxima were almost the same for $\xi=1$ and $\xi=2$. 

\paragraph{Computing $p$-values of the local score}

Assuming that the values of the scores are independent, formulas for computing the $p$-values 
of the global maximum of the Lindley process have been proposed in different 
scenarios (the main difference was the length of the sequence and the computation of
approximations or exact $p$-values)  \citep{KAR1990, KAR1992, MER2001, MER2003}. 
Since then,
local score procedures have been used, for example for aligning sequences, and the independence assumption
was always used for computing the $p$-values, even if it was clearly wrong. This shows how difficult it is to 
take dependence into account. 
Following ideas of \citep{ROB2005} we proposed to find the empirical distribution of the local score, 
as a function of $\rho$ and $L$, for a particular $\xi$. 
From simulations, we learned how
the parameters $a-log(L)$ and $b$ depend from $\rho$ (for a fixed $\xi$), given that the distribution of the 
$p$-values of single marker tests is uniform. 
The polynomes given in Supporting Information can now be used to compute parameters $a$ and $b$ of the Gumbel 
distribution in future studies, 
(even related to a totally different question),
as long as single marker $p$-values are uniformly distributed under the null distribution, and $\xi=1$ or 2.

If the $p$-values do not follow a Gumbel distribution, a re-sampling strategy, similar to the one implemented for 
the Quail data, will be necessary.
In the quail data, this strategy could be implemented because we observed from simulations that the parameters 
$a-log(L)$ and $b$  converged to a limit for not too large values of $L$.
This allowed to estimate parameters of long sequences using simulations from shorter sequences.
In the future, further investigations should however be carried out to find a better and more general way of 
estimating the $p$-values.

Our method supposes a constant auto-correlation along the each chromosome. Even if we know that this is not true, 
this provided at 
least a way to take marker dependencies into account. 
Further research is also needed in order to consider changes of LD along the chromosome, which would likely require 
a modification 
of the score function.

\paragraph{Pros and cons of competing detection methods in presence of high drift}

The quail experiment perfectly illustrates the difficulty to distinguish selective processes from 
neutral processes, and the importance to cumulate signals from multiple markers
to overcome it. Indeed, a lot of drift has been cumulated in the two quail lines, 
because only 60 individuals were kept at each generation,
so that markers with high $F_{ST}$ values are very likely even under the null hypothesis of neutrality.
For instance, $1.5 \%$ of the markers on GGA1 were fixed differentially in the two lines and thus 
achieved the maximal $F_{ST}$ value. A large proportion of these differential fixations might be just due 
to drift.
Consequently, considering all markers in the top $1\%$ of the $F_{ST}$ distribution as selection 
candidates, which is a common practice in single marker based genome scans,
must result here in a large proportion of false positives.

As selection does not only raise up the frequency of the selected allele, but also of other alleles 
in its neighborhood, we expect to find an increased amount of markers with high $F_{ST}$ value around a 
selected mutation. \cite{JOH2010} exploited this property by looking for 
clusters of alleles that were fixed for alternative alleles in the divergent lines.
Because of linkage disequilibrium, we expect that some of the alleles in the neighborhood of 
the selected variant will be fixed differentially in the two lines.
However, in this type of experiments, selection generally does not act on new variants 
(divergence time is probably too small for new advantageous mutations to appear)
but on variants that were already segregating in the founder population. So, at many other sites 
we just expect an increased allele frequency difference between the two lines, 
but not necessarily differential fixation.
Consequently, alleles showing a large difference in allele frequencies between the two lines are also 
informative, even though a bit less than differentially fixed ones. 

The local score approach tries to take advantage of this information, as alleles that are not 
fixed but have a high $F_{ST}$ value also contribute to the local score. 
Actually, the local score proposed here can be seen as a generalization of the clustering method of 
\cite{JOH2010}, where the markers would
get a positive score if they are differentially fixed, and a negative or zero score otherwise.
When analyzing the quail data, we tried a windowing approach based on the number of differentially 
fixed SNPs in each window.
Although our definition of fixed SNPs was a bit more liberal than that of \citep{JOH2010} (the allele 
had to be fixed in only one line), this approach was quite similar in spirit. 
No clear signal could be detected with this approach, in contrast to the local score, confirming the 
fact that cumulating information from all allele frequency differences is very important.

Another interest of cumulating signals from consecutive markers, as implemented in the local score, is to reduce the noise 
arising from sequencing errors.
Such errors are unavoidable with next bio-technologies in general, and with next generation sequencing in particular.
They result in a lower accuracy when estimating allele frequencies at a SNP, 
which is even decreased in the case of pooled samples with small sample size.
In the quail experiment, pools of only 10 diploid individuals were used.
But, using a method specifically designed to improve the estimation of allele frequencies from pooled samples \citep{BOI2013} 
(see also \citep{Ferreti2013} who showed that individual and pooled estimates can be highly correlated),
and cumulating single marker signals using the local score, we could extract rich information from this relatively small sample 
size.

\paragraph{Looking for information from founder population}
Sequencing the founder population of the two quail lines would increase our power to detect selection and improve our 
understanding of the selective process in a given region \citep{LAV2003a}.
Indeed, sequencing the ancestral population provides a precise estimation of founder allele frequencies. 
Without this information, founder allele frequencies are estimated by the mean of the allele frequencies in the two divergent 
lines. This estimation procedure is the best we can imagine in this situation, but it is clearly biased for sites under selection.
Imagine for instance that an allele with initial frequency 0.1 increases in frequency in line 1 due to selection, and is lost in 
line 2 due to drift. The allele frequency trajectory in line 1 clearly suggests the influence of positive selection.
But from the final allele frequencies (0 and 1) we will assume that founder allele frequency was 0.5, so that allele frequency 
trajectories in the two populations are less informative. 
Similarly, the identification of the SNPs and the population with the highest allele frequency variation would be much more 
accurate if ancestral allele frequencies were known.

For future similar studies, we therefore recommend to also sequence individuals from the founder population or, if  not 
possible, to sequence a higher number of present populations in order to better estimate ancestral allele 
frequencies (see the discussion in \citep{Maxime, FAR2013}).

\section{Conclusion}  
\label{sec:Conclusion}

This work enhanced the added value of a divergent selection experiment on a behavior trait, 
pool sequencing of two divergent lines, and an appropriate statistical approach: here the local score. 
All this combined lead to the discovery of eight small genomic regions exhibiting  candidate genes related to autism and 
social behavior.

Besides, other studies, like Genome Wide Association Studies (GWAS), could also gain in applying this local score strategy, 
as proposed by \cite{GUE2006}, \cite{ROB2005}, \cite{TEY2012},
in particular when $p$-values are not extreme and when marker density is high (as in the experiment of \cite{JOH2010}). 
In genome scans for selection, the local score approach can be applied to data with or without individual genotypic information
and any statistic.

\singlespace
\nolinenumbers

\section{Supplementary Material}
Supplementary tables S1-S7 and figures S1-S11 are available  at 

\section{Acknowledgments}
This work was supported by the French Agence Nationale de la Recherche provided to C.L. (SNP-BB project, ANR-009-GENM-008) for the sequencing,
and to M.S.C. and S.B. (D\'eLiSus project, ANR-07-GANI-001) for the methodology. 
The R\'egion Midi-Pyr\'en\'ees, the D\'epartement de G\'en\'etique Animale of Institut National de la Recherche Agronomique
 and Agencia Nacional de Investigaci\'on e Innovaci\'on provided financial support for the salary of M.I.F.
Sequencing was performed at GeT-PlaGe Genotoul platform.
Data analyses were performed on the computer cluster of the bioinformatics platform Toulouse Midi-Pyr{\'e}n{\'e}es. 
We thank Bertrand Servin for stimulating discussions.


\bibliography{hapflk_scorelocal}

\newpage
\onecolumn

\vspace*{5cm}

\newpage

\newpage

\vspace*{1cm}

\newpage
\vspace*{3cm}

\begin{table}[!h]
\begin{tabular}{cccccl}
\small
  Reg. & GGA & Position &  L (kb) & Gene name & Description \\
 \hline
  1 & 1 & 92,963,200-93,182,431 &  219 & NSUN3  & putative methyltransferase NSUN3\\
  &   &    &                & ARL13B & ADP-ribosylation factor-like protein 13B \\
 2 &  2 & 1,583,808-1,688,282 & 105 & VIPR1  & vasoactive intestinal polypeptide receptor\\
   &   &                     &     &        & 1  precursor\\
 3 & 3 & 61,585,151-61,604,462 & 19 & ECHDC1 & ethylmalonyl-CoA decarboxylase\\
   &     &   &                & RNF146 & ring finger protein 146\\
 4 &  3 & 75,088,236-75,170,475 & 82 & MMS22L & protein MMS22-like\\
 5 & 4 & 11,412,108-11,452,505 & 40 & GLOD5  & Glyoxalase domain-containing protein 5\\
 6 &  4 & 90,952,530-91,008,189 & 56 & CTNNA2 & catenin (cadherin-associated protein) $\alpha$-2 \\
 7 & 6 & 35,234,541-35,336,717 & 102 & FOXI2  & forkhead box protein I3 \\
   &  &     &                 & PTPRE  & receptor-type tyrosine-protein phosphatase\\
   &  &     &                 &        & $\epsilon$ precursor\\
 8 & 6 & 6,311,686-6,644,350 & 333  & UBE2D1 & ubiquitin-conjugating enzyme E2 D1\\
   &  &     &                  & CISD1  & CDGSH iron sulfur dom-containing prot 1\\
   &  &     &                 & IPMK   & inositol polyphosphate multikinase\\

\end{tabular}
\caption[Footprints of selection in quail]{Footprints of selection in the quail experiment including two divergent 
lines for social behavior, and the genes
included in each detected region (mapped on chicken genome).}
\label{tab:caille_liste_regions}
\end{table}

\end{document}